X-Authentication-Warning: thwgs01.cern.ch: amperrin owned process doing -bs
Date: Thu, 20 Jul 2000 15:29:06 +0200 (MET DST)
From: Anne-Marie Perrin <Anne-Marie.Perrin@cern.ch>
X-Sender: amperrin@thwgs01
To: Alvaro De Rujula <derujula@nxth04.cern.ch>

\documentstyle[12pt]{article}
\begin{document}



\pagestyle{empty}
\begin{flushright}
{CERN-TH/2000-216}\\
\end{flushright}
\vspace*{5mm}
\begin{center}
{\bf WHAT IS THE COSMIC-RAY LUMINOSITY OF OUR GALAXY?} \\
\vspace*{1cm}
{\bf Arnon Dar} \\
Technion, Israel Institute of Technology, Haifa, Israel\\
and \\
Theoretical Physics Division, CERN \\
CH - 1211 Geneva 23 \\
\vspace*{0.5cm}
and\\
\vspace*{0.5cm}
{\bf A. De R\'ujula}\\
Theoretical Physics Division, CERN \\
CH - 1211 Geneva 23 \\
\vspace*{2cm}
{\bf ABSTRACT} \\ \end{center}
\vspace*{2mm}
\noindent
The total cosmic-ray luminosity of the Galaxy is an
important constraint on models of cosmic-ray generation.
The diffuse high energy $\gamma$-ray and radio-synchrotron
emissions of the Milky Way are used to derive this
luminosity. The result is almost two orders of magnitude larger
than the standard estimate, based on the observed isotopic abundances of 
cosmic ray nuclides. We discuss the plausible interpretation of
this discrepancy and the possible origin of such a relatively large
luminosity.

\vspace*{5cm}

\begin{flushleft} CERN-TH/2000-216 \\
July 2000
\end{flushleft}
\vfill\eject

\setcounter{page}{1}
\pagestyle{plain}




\setlength{\parskip}{0.45cm}
\setlength{\baselineskip}{0.75cm}


Almost a century after they were discovered, our understanding
of cosmic rays is still very limited. Their production mechanisms,
composition and energy spectrum continue to be debatable.
In this letter we discuss the total cosmic ray (CR) luminosity
of our Galaxy,
a crucial constraint on models of galactic CR generation.

The CR nuclei have a power-law
spectral flux ${\rm dF/dE\propto E^{-\beta_i}}$
with a series of break-point
energies: $\beta_{1,2,3}$ $\sim 2.7,~3.0$, and 2.5 in the intervals
${\rm  10~GeV < E < E_{knee}}$ $\sim 3 \times 10^{6}$ GeV,
$~{\rm E_{knee} < E < E_{ankle}}$ $\sim 3 \times 10^{9}$ GeV,
and
${\rm E_{ankle} < E < }$ $3\times 10^{11}$ GeV.
Below ${\rm E_{knee}}$, protons constitute  $\sim 96\%$
of the CRs at fixed energy per nucleon. Their flux and number density
above ${\rm E_p\sim 10}$ GeV is (see, for instance, Wiebel-Sooth and
Biermann
1998 and references therein):
 \begin{equation}
{\rm {dF_p\over dE}\simeq 1.8 \left
[E \over GeV \right]^{-2.70\pm 0.05}
~cm^{-2}~s^{-1}~sr^{-1}~GeV^{-1}}.
\label{protons}
\end{equation}
The spectral indices of heavier nuclei are compatible within errors
with that of protons and, given the dominant abundance of the latter,
we need not distinguish here between CR protons and the ensemble of
nuclear CRs. It is generally
believed that the bulk of the CR nuclei with energy below the knee are
Galactic in origin, and that their main production mechanism is
acceleration by supernova shocks (see, for instance, Ginzburg and Syrovatskii
1964; Longair 1981; Berezinskii et al. 1990; Gaiser  1990).

If the CRs are chiefly Galactic in origin, their accelerators must
compensate for the escape of CRs from the Galaxy, in order to sustain the
observed Galactic CR intensity: it is known from meteorite records that
the CR flux has been steady for the past few giga-years (Longair 1981).
The Milky Way's luminosity in CRs must therefore satisfy:
\begin{equation}
{\rm L_{CR} \simeq L_{p}={4\pi \over c} \int {1\over \tau_{conf}}
\,E\,{dF_p\over dE}\,dE\,dV}\, ,
\label{CRlum}
\end{equation}
where ${\rm \tau_{conf}(E)}$ is the mean confinement time in the Galaxy
 of CRs of energy E.

The standard estimate of ${\rm L_{CR}}$ runs along the following lines.
The mean column density X traversed by CRs before they reach the Earth
can be extracted from the observed ratios of primary to secondary CRs.
The result is ${\rm X\approx 6.9\,[E(GeV)/(20\, Z)]^{-0.5}~g~cm^{-2}}$
(Swordy et al. 1990). With
use of ${\rm X=\int \rho\, dx\sim\bar\rho\, c\, \tau_{conf}}$ one can
extract the product of ${\rm \tau_{conf}(E)}$ and a path-averaged
density $\bar\rho$.
Assume the locally measured values of X and ${\rm dF_p/dE}$
to be representative of the Galactic values dominating the integral in  
Eq.~(\ref{CRlum}),
to obtain:
\begin{equation}
{\rm L_{CR}\sim {4\pi \over c}
\int \bar\rho \,dV\int {1\over X}\,E\, {dF_p\over dE}\,dE}\, .
\label{them0}
\end{equation}
Assume the path-averaged $\bar\rho$ to be close to the average
density $\rho$ of neutral and ionized gas in the Galaxy, so that
${\rm \int\bar\rho\, dV}$ is the total mass of Galactic gas,
estimated from X-ray, optical and radio observations (Longair 1981) to be
${\rm M_{gas}\approx 4.8\times 10^9\, M_\odot}$.
The integration over energy is not unduly sensitive to its lower limit
and converges rapidly above the knee.
The final result (Drury et al. 1989) is:
\begin{equation}
{\rm L_{CR} \sim 1.5\times 10^{41}~erg~s^{-1}}\, .
\label{them}
\end{equation}
Earlier estimates (e.g. Berezinskii et al. 1990 and references therein) 
of ${\rm L_{CR}}$, which used the ``Leaky Box''
 model of CR confinement, led to somewhat smaller luminosities.

In spite of the cursory character of the above luminosity estimate,
Eq.~(\ref{them}) is consistent  with the
assumption that CRs are dominantly accelerated by
the turbulent magnetic fields of supernova (SN) remnants,
generated by the expansion of the debris from the SN explosion
into the interstellar medium. For an estimated mean Galactic rate of one
supernova every $\sim$ 50 years (van den Bergh and Tammann 1991)
 and an average kinetic energy
${\rm \langle E_k\rangle\approx 10^{51}~erg}$ of the debris, this explanation
requires an  $\epsilon \sim 20\%$ efficiency in the conversion of kinetic
energy into CRs.

The model of CR generation by SNe is incomplete or
problematic in several respects (e.g. Plaga et al. 1999). Supernova-generated
shocks are not sufficiently lasting and energetic to produce
CRs with energies well above ${\rm E_{knee}}$
(e.g. Lagage and Cesarsky 1983). The space
distribution of SNe is too concentrated in the Galactic disk
and bulge to give a proper description of the relative isotopic
abundances of CRs, in particular ${\rm ^{10}Be/^9Be}$
(Strong and Moskalenko 1988),
of the directional distribution of the diffuse $\gamma$-ray
background radiation (Strong and Mattox, 1996)
 and of the high energy $\gamma$-rays
produced by CR interactions in the interstellar medium
(Strong and Moskalenko 1998).
The CR luminosity of SN remnants is severely constrained
by TeV $\gamma$-ray observations and results in a
 CR-generation efficiency $\epsilon$ between 1 and $5\%$
(Allen et al. 1999),
somewhat short of the required $\epsilon\sim 20\%$.


In this letter we present an alternative estimate of the
luminosity of CR nuclei, based on the Galactic CR-electron
luminosity, which we infer from observations
of  $\gamma$-ray production (Hunter et al. 1997; Sreekumar et al. 1998)
 and synchrotron emission (e.g. Chen et al. 1996)
by CR electrons.

The EGRET detector on the Compton GRO satellite has mapped the
intensity and spectral index of the ``diffuse'' $\gamma$-ray
background (GBR) above ${\rm E_\gamma=30}$ MeV,
at latitudes above the Galactic disk and bulge. The observed spectrum, 
${\rm dF/dE_\gamma\propto E^{-\beta_\gamma}}$,
has an index $\beta_\gamma\simeq 2.10\pm 0.03$ that is independent
of direction. The intensity is also roughly isotropic, thus the claim
of a dominantly extragalactic origin of the GBR.

We have recently shown (Dar and De R\'ujula, 2000)
 that the GBR intensity is significantly
correlated with the angle away from the galactic centre and that
it is dominated at high latitudes  by inverse Compton scattering (ICS) of
CR electrons from the cosmic microwave background radiation (CBR)
and from starlight,  obviating the recourse to unspecified
extragalactic sources (the importance of ICS has also been stressed
in ,  Strong and Moskalenko 1998; Strong et al. 2000).
Earlier evidence for a large galactic
contribution to the GBR at large latitudes had been found
by Chen et al.~(1996), who discovered a strong
correlation between the observed EGRET GBR $\gamma$-ray intensity and
the galactic radio continuum emission at 408 MHz, which is dominated by
synchrotron radiation from the very same CR electrons that produce
$\sim$ 100 MeV $\gamma$-rays by ICS from galactic stellar light.

Our model of the origin of the GBR is based on the assumption that
the average Galactic CR-electron spectrum has the same
energy dependence as the locally observed one (for a recent
compilation of experimental results see Wiebel-Sooth and Biermann 1998).  
This spectrum is well fit, from $\sim 10$ GeV to $\sim 2$ TeV, by:
\begin{equation}
{\rm {dF_e\over dE}\simeq (2.5\pm 0.5)\times 10^5 \left
[E \over MeV \right]^{-3.2\pm 0.10}
~cm^{-2}~s^{-1}~sr^{-1}~MeV^{-1}}.
\label{electrons}
\end{equation}
Starlight and CMB photons, upscattered by electrons with the spectral
index of  Eq.~(\ref{electrons}), have an energy dependence with index  
$\beta_\gamma=(\beta_e+1)/2= 2.10\pm 0.05$, in perfect agreement
with the observed direction-independent index of the GBR
(Dar et al. 1999, Dar and De R\'ujula, 2000).

The relation between the spectral indices of CR protons and CR electrons
in Eqs.~(\ref{protons}) and (\ref{electrons}) can also be understood
in very simple
terms. The confinement, residence or accumulation
time of nuclear CRs in the Galaxy is $\rm \tau_{conf}(E)\propto$
${\rm X\propto E^{-0.5}}$. The result must be of the same form
for electrons, since relativistic particles of the same charge behave in the
same way in a magnetic maze.
The source spectrum of nuclear CRs,
$\rm dF_p^s/dE$, is related to the observed spectrum of CR nuclei
by $\rm dF_p/dE\propto \tau_{conf}(E)\,dF_p^s/dE$, so that the index of
$\rm dF_p^s$ is $\rm \beta^s=\beta_p-0.5\simeq 2.2$.
If the mechanism accelerating CR hadrons and
CR electrons is the same e.g., first-order acceleration
by a moving magnetic field (Fermi 1949, 1954)
$\rm dF_p/dE\propto dF_e^s/dE$,
and the source spectral index for electrons is also $\rm \beta^s\simeq 2.2$.
Electrons, unlike nuclei, are significantly affected by ICS and synchrotron
cooling --whose characteristic time is $\rm \tau_{cool}\propto m^2/E$--
so that, at sufficiently high energy,  cooling takes over the
accumulation-time effect in modulating the electron spectrum.
The result (Dar et al. 1999, Dar and De R\'ujula, 2000) is
$\rm \beta_e=\beta^s+1\simeq 3.2$, in agreement with Eq.~(\ref{electrons}).

In our study of the GBR we adopted, for
the spatial distribution of the CR electron flux in the Galaxy,  a
model with a gaussian scale height ${\rm h_e}$ above the Galactic
plane, and a scale radius  ${\rm \rho_e}$ in directions perpendicular
to the Galactic axis. By adjusting ${\rm h_e\sim 20}$ kpc and
${\rm \rho_e\sim 35}$ kpc, we reproduced the observed intensity and
angular dependence of the GBR.
We shall assume the nuclear CRs
to be distributed as the CR electrons, with the ratio of fluxes fixed at
its locally observed value. A scale height of CR nuclei
${\rm h_{CR}= h_e\sim}$ 20 kpc is larger than conventionally assumed,
but is not excluded by data on relative CR abundances. For the most
elaborate models (Strong and Moskalenko 1998)
 a ``Leaky-Box'' scale height of 20 kpc is only some 1.3 standard
deviations below the central value of the most precise observations
(Connell 1998) and is perfectly compatible with the
average of all previous and somewhat less precise
results, compiled in Lukasiak et al.~(1994).
Since our CR distribution
is much more extensive than the visible part of the Galaxy, we refer
to it as the ``cosmic-ray halo''.

The EGRET GBR data to which we fit the properties of a CR electron halo
are gathered by masking the galactic plane at latitudes
$\rm{|b|\le 10^o}$, as well as the galactic centre
at $\rm{|b|\le 30^o}$ for longitudes $\rm{|l|\le 40^o}$. The volume of
the CR halo is so much larger than that of the Galaxy within EGRET's
mask, that it is a very good approximation --in computing the Galaxy's
total CR luminosity-- to use our halo model throughout the entire
Galaxy (within the mask the model accounts for $\sim 1/2$ of the
observed diffuse $\gamma$ radiation).

The successful relation between the spectral indices of the GBR
and the CR electrons followed from the assumption that the production rate of
CR electrons is equal to their cooling rate, which we estimate as follows.
The  starlight-photon density in the CR halo
may be obtained by approximating our galaxy's
starlight as that produced by a source at its centre with the galactic
luminosity ${\rm L_\star}=2.3\times 10^{10}$ ${\rm L_{_\odot}}$
$\simeq 5.5~10^{55}$ eV s$^{-1}$ (Pritchet and van den Bergh 1999):
${\rm n_\star\approx L_\star / (4\,\pi\,c\,\epsilon_\star\,r^2)}$,
where $\epsilon_\star\sim 1$ eV is the average photon energy.
For a gaussian CR halo, the mean ${\rm n_\star}$
 is given by:
\begin{equation}
{\rm \langle n_\star\rangle \approx {L_\star\,\over
4\,\pi\,c\,\epsilon_\star\,\rho_e^2}\,{1\over u}\, ln \left({1+u\over 1-u}
\right)\approx 0.035\,cm^{-3}}\, ,
\label{MeanLight}
\end{equation}
with ${\rm u^2=1-h_e^2/\rho_e^2}$.
The mean energy density of starlight in the CR halo is
much smaller than that of the CBR (${\rm \sim
0.24~eV~cm^{-3}}$). If the local magnetic field-energy is in
equipartition with the CR energy density, ${\rm B^2/8\pi\sim
1~eV~cm^{-3}}$, the mean total electromagnetic energy density in the
halo is ${\rm \rho_\gamma \approx 1.27~eV~cm^{-3}}$.
For the electron energy range of  interest
the Thomson limit of the $e\gamma$ cross section
(${\rm \sigma_{_T}\approx 0.65\times 10^{-24}~cm^2}$) is accurate,
even for ICS on starlight, and
the mean cooling rate, ${\rm R_c}$, of CR electrons by ICS and
synchrotron radiation is (Dar and De R\'ujula 2000):
\begin{equation}
{\rm R_c(E)\equiv{1\over \tau_{cool}(E)}\approx {4\,\rho_\gamma\,
\sigma_{_T}\, c\,E \over 3\,(m_e\, c^2)^2}
\approx 4.0 \, \left [{E\over GeV}\right]~Gy^{-1}}\, .
\label{coolingrate}
\end{equation}
The luminosity of our galaxy in high energy electrons of energy above E,
in  equilibrium with their cooling rate by ICS and synchrotron radiation, is:
\begin{equation}
{\rm L_e (>E) \approx h_e\,\rho_e^2 \,
{4\,\pi^{5\over 2}\over c}\int dE\,E\,{\partial\over\partial E}
\left(R_c\,{dF_e\over dE}\right)}\,.
\label{elum}
\end{equation}
By substituting the flux of Eq.~(\ref{electrons}), we obtain:
\begin{equation}
{\rm L_e(>E) \approx 1.13\times 10^{41}
\left[{h_e\over 20~kpc}\right]
\left[{\rho_e\over 35~kpc}\right]^2
\left[{E\over~2.5\,GeV}\right]^{-0.20\pm 0.10}~erg~s^{-1}}\, .
\label{eluminosity}
\end{equation}

We have assumed that the ratio of the CR nuclear and electron fluxes
is universal throughout the Galaxy. Thus, to
estimate the Galaxy's CR luminosity, it suffices to scale the electron
luminosity ${\rm L_e}$ by the local ratio R of CR and electron fluxes.
For the total fluxes, this ratio is $\rm R\sim 80$. This result
is uncertain, since it is dominated by
CR-energies of $\cal{O}$(1) GeV, a domain in which the fluxes are affected
by local magnetic and solar-wind effects.
An independent estimate of R can be obtained as follows. The CR electron
spectrum sharply steepens to an index $\rm \beta_e\simeq 3.2$ at
$\rm E\sim 5$ GeV: that must be the energy at which ICS and synchrotron 
radiation, for which $\rm \tau_{cool}\propto E^{-1}$, take over the effect of
CR accumulation, for which $\rm \tau_{conf}\propto E^{-0.5}$
(the effects of local magnetic fields, the solar wind, Coulomb scattering,
ionization losses and bremsstrahlung are only relevant at even lower
energies).
Thus, to within a factor of $\cal O$(2), the observed proton to electron
flux ratio at $\rm E=5$ GeV must be the ratio of their source fluxes.
We have argued that the source fluxes have the same spectral index.
Thus, their ratio at fixed energy is also their energy-integrated ratio.
This gives ${\rm R\sim 60}$, in rough agreement with the previous estimate.

Multiply $\rm L_e$ in Eq.~(\ref{eluminosity}) by
$\rm R=60$, to obtain:
\begin{equation}
{\rm L_p(>E) \sim 6.8 \times 10^{42}
\left[{h_e\over 20~kpc}\right]
\left[{\rho_e\over 35~kpc}\right]^2
\left[{E\over~2.5\,GeV}\right]^{-0.20\pm 0.10}~erg~s^{-1}}\, ,
\label{pluminosity}
\end{equation}
which is almost two orders of magnitude larger than the
estimate of Eq.~(\ref{them}).


If, as we argued, $\rm \tau_{cool}=\tau_{conf}$ at $\rm E\sim 5$ GeV,
we can use Eq.~(\ref{coolingrate}) to obtain
$\rm \tau_{conf}= 250$ My at $\rm E=1$ GeV. This is an
order of magnitude larger than the values of $\rm \tau_{conf}$ obtained
from the analysis of the relative abundances of unstable to
stable CRs: $\rm ^{10}Be/Be$ (Lukasiak  et al. 1994;  Connell 1998),
 $\rm ^{26}Al/^{27}Al$  (Lukasiak et al. 1994b;
Simpson and Connell 1998) and $\rm ^{36}Cl/Cl$ (Connell et al. 1998).
The discrepancy is even larger, since these data are for lower energies,
 of $\cal O$(250) MeV per nucleon.
This alterity can be easily understood (Plaga 1998). The confinement
time is extracted from the isotopic ratios using  a Leaky Box
model, wherein the magnetic field of the Galaxy is confined to a
region of dimensions similar to those of the visible part of the Galaxy. 
But, if the dense and luminous component of the Galaxy is embedded
--as we surmise-- in a much
larger and less dense magnetized halo, the stable CRs may spend
much of their travel time in the halo, while the unstable ones must
have much shorter trajectories (the lifetimes of $\rm ^{10}Be$,
$\rm ^{26}Al$ and $\rm ^{36}Cl$
are a mere 1.6, 0.87, and 0.30 My, respectively). This is also
the reason why our estimate of the CR luminosity, Eq.~(\ref{pluminosity}),
is not truly contradictory to the much smaller conventional estimate
of Eq.~(\ref{them}): the gas density, volume and grammage used to
derive Eq.~(\ref{them}) all refer to CRs confined to a region close
to the visible Galaxy.

What CR acceleration mechanism could give rise to the large
luminosity of Eq.~(\ref{pluminosity})? The bulk of the
high energy CRs may be accelerated by relativistic jets emitted in
the birth of neutron stars and stellar black holes in supernova
explosions (Dar and Plaga 1999).
If the mean sky velocity of neutron stars (Lyne and Lorimer,1964),
${\rm \langle v_{ns}\rangle\simeq
450\pm 90~km~s^{-1}}$, is due to an imbalance in this relativistic jet
ejection,  ${\rm E_{jet}>M_{ns}\, v_{ns}\, c\simeq 4\times
10^{51}~erg}$, for $\rm M_{ns}=1.4\, M_\odot$. If the kinetic energy
of the jets is efficiently
converted to CR energy, and  for the estimated rate
(van den Bergh and Tammann 1991)
of Type II, Ib and Ic supernovae, $\rm R_{SN}\sim  1/50$ per year, then:
\begin{equation}
{\rm L_{CR}\simeq 2\, E_{jet}\, R_{SN}
\approx 5.1 \times 10^{42}~erg~s^{-1}}\, ,
\label{crluminosity}
\end{equation}
in reasonable agreement with Eq.~(\ref{pluminosity}).

Relativistic jets are emitted by active galactic nuclei
and by galactic microquasars. These jets are observed
to consist of ``plasmoids'' whose cross section,
after an initial period of transverse expansion at the
speed of sound in a relativistic plasma ($\rm c/\sqrt{3}$),
remains surprisingly constant until the jet sweeps
enough material to stop and disperse as a blob
(Rodriguez and Meribel 1999). If the plasmoids of the
jets allegedly responsible for the peculiar velocities of neutron
stars have a Lorentz factor $\rm \gamma=E/M\,c^2$ of $\cal{O}$(10$^3$),  
they are
good candidate sources of $\gamma$-ray bursts
(Dar and Plaga 1999). If the transverse size
of these plasmoids of $\cal{O}$(0.1) pc, the column density necessary
to stop them is of the same order as the one transverse to the
galactic disk. Thus, the jets may reach the halo of the Galaxy
before they stop, seeding it with a CR population and a magnetic
field, and giving consistency to our overall picture.

\section*{REFERENCES}

\begin{description}
\item
Allen, G.E., Gotthelf, E.V. \& Petre, R. {\it Proceedings of the  
26$^{th}$ International Cosmic Ray Conference} (Eds. Kieda, D., Salamon, M.,
\& Dingus, B. Salt Lake City, Utah, 2000).
\item
Berezinskii, V. S. et al. 1990, {\it Astrophysics
of cosmic rays} (North Holland, Amsterdam).
\item
Chen, A. Dwyer, J. \& Kaaret, P., 1996, ApJ, 463, 169.
\item
Connell, J. J. et al. 1998, ApJ, 509, L97.
\item
Connell, J. J. 1998, ApJ, 501, L59.
\item
Dar, A. \& Plaga, R., 1999, A\&A, 349, 257 (1999).
\item
Dar, A., De R\'ujula, A. \& Antoniou, N.,  astro-ph/9901005 (Proc. Vulcano
Workshop 2000, in press).
\item
Dar, A. \& De R\'ujula, A., 2000, preprint astro-ph/0005080.
Drury, L. O'C., Markiewicz, W. J. \&  V\"olk,  H. J., 1989, A\&A, 225, 179.
\item
Fermi, E., 1949,  Phys. Rev. 75, 1169.
\item
Fermi, E., 1954,  ApJ.  119, 1.
\item
Gaiser, T. K., 1990 {\it Cosmic rays and particle physics}
(Cambridge Univ. Press).
\item
Ginzburg, V. L. \&  S. I. Syrovatskii, S. I. 1964, {\it  The origin of
cosmic rays} (Pergamon Press, Oxford).
\item
Hunter, S. D.  et al. 1997, ApJ. 481, 205.
\item
Lagage, P. O. \& Cesarsky, C. J. 1983,  A\&A, 125, 249.
\item
Longair, M. S.  1981, {\it High energy astrophysics} (Cambridge Univ.
Press).
\item
Lukasiak A., McDonald F.B. \& Webber W.R., 1994
AJ, 423 (1994) 426.
\item
Lukasiak, A. et al., 1994b, ApJ, 430, 69L.
\item
Lyne, A. G., \& Lorimer, D. R., 1964, Nature, 369, 127.
\item
Moskalenko, I. V. \& Strong A. W., 2000, ApJ, 528, 327.
\item
Plaga. R, 1998, A\&A, 330, 833.
\item
Plaga. R, de Jager, O. C. \& Dar, A., 1999,  astro-ph 9907419.
\item
Pritchet C. J., \& van den Bergh S., 1999, AJ, 118, 833.
\item
Rodriguez L.F., \& Mirabel, I.F., 1999, ApJ, 511, 398.
Simpson, J.A. \& Connell, J.J., 1998, ApJ, 497, L85.
\item
Sreekumar, P.  et al., 1998,  ApJ. 1998, 494, 523.
\item
Strong, A. \& Moskalenko, I. V., 1998, ApJ, 509, 212.
\item
Strong, A. \& Moskalenko, I. V., and Reiner, O., 2000, ApJ 537 (in press).
\item
Strong, A. W. \& Mattox, R. J., 1996 A\&A, 308, L21.
\item
Swordy, S. P. et al., 1990, ApJ 330, 625.
\item
van den Bergh, S. \& Tammann, G. A., 1991, ARA\&A,  29, 363.
\item
Wiebel-Sooth, B. \& Biermann, P. L., 1998, {\it Cosmic Rays}
(Landolt-Bornstein; Springer Verlag,
Heidelberg, 1998).


\end{description}

\end{document}